\documentclass[journal=ancac3,manuscript=article]{achemso}
\usepackage{amsmath,amssymb,bm}
\usepackage{graphicx}
\usepackage{siunitx}
\usepackage[version=4]{mhchem}
\usepackage{xcolor}

\author{Zhijian Zhou}
\altaffiliation{These authors contributed equally.}
\affiliation{Department of Chemical and Biomolecular Engineering, University of Pennsylvania, Philadelphia, Pennsylvania 19104, United States}
\author{Claire Dor\'e}
\altaffiliation{These authors contributed equally.}
\affiliation{Department of Chemical and Biomolecular Engineering, University of Pennsylvania, Philadelphia, Pennsylvania 19104, United States}
\author{Yusheng Cai}
\affiliation{Department of Chemical and Biomolecular Engineering, University of Pennsylvania, Philadelphia, Pennsylvania 19104, United States}
\author{Jed-Joan Edziah}
\affiliation{Department of Chemical and Biomolecular Engineering, University of Pennsylvania, Philadelphia, Pennsylvania 19104, United States}
\author{Brian Edwards}
\affiliation{Department of Electrical and Systems Engineering, University of Pennsylvania, Philadelphia, Pennsylvania 19104, United States}
\author{Nader Engheta}
\affiliation{Department of Electrical and Systems Engineering, University of Pennsylvania, Philadelphia, Pennsylvania 19104, United States}
\author{Amish J. Patel}
\affiliation{Department of Chemical and Biomolecular Engineering, University of Pennsylvania, Philadelphia, Pennsylvania 19104, United States}
\author{Kathleen J. Stebe}
\email{kstebe@seas.upenn.edu}
\affiliation{Department of Chemical and Biomolecular Engineering, University of Pennsylvania, Philadelphia, Pennsylvania 19104, United States}

\title{Flexoelectric Switching of Lock-and-Key Topological Bits}


\begin{document}


\begin{abstract}
Topological defects in nematic liquid crystals are high-energy sites of reduced order. We show that a point defect can be stabilized in pillbox-shaped pores. The pore shape promotes lock-and-key docking similar to that observed for colloidal particles. The defect docks in one lobe or the other, displaying bistability. Furthermore, the defect moves like a charged object in an external electric field. An in-plane field drives the defect from one lobe to the opposite lobe, where it remains once the field is removed. The velocity is linear in field strength and reverses with polarity, consistent with flexoelectric coupling. Within the pore, the nematic around the defect carries a positive bulk bound charge density, compensated by negative space charge in the rest of the fluid and on the bounding surfaces. The external field breaks the mirror symmetry of the pore and drives the defect motion. Defect displacement increases the component of the net dipole in the domain along the field. Antagonistic curvature in peanut-shaped pores repels the defect, allowing pore geometry to set distinct switching thresholds. The defects, while not topologically required, are long lived, so information can be stored in their position, allowing them to serve as topological bits. Arrays of topological bits store and rewrite binary information in defect positions, offering a route to reconfigurable, field-addressable soft materials.
\end{abstract}

\section{Introduction}

A wide variety of topological defects self-assemble in nematic liquid crystals, including point defects, disclination lines, loops, knots and links \cite{Kleman2006,Tkalec2011}, and umbilics \cite{KimSerra2020}.
Defects generate optical textures defined by the liquid crystal's birefringence and the spatial arrangement of the local director field around them. Their spontaneous formation, optical signatures, topological protection and ability to be restructured make them exciting vehicles for reconfigurable optical elements \cite{KimSerra2020,Ma2022}.

Defects can be created, positioned, and transformed. Patterned and curved surfaces pin defects and influence defect structure \cite{Tran2016,Li2017,OhzonoZLDA,Kim2018Mosaics}. Laser tweezers cut and rejoin loops to form knotted structures \cite{Tkalec2011}. Electric fields act on defects through dielectric reorientation of the surrounding director \cite{Harkai2020} or through flexoelectric coupling \cite{Pieranski2019,PieranskiEPJE2019,Pieranski2020,BryanBrownGABN,Spencer2010}.

In patterned cells, some defect arrays form only under applied fields and return to a ground state once the field is removed \cite{KimSerra2020}. Other defect systems are stabilized by confinement, e.g. in micropost arrays \cite{Kara2025} and cubic microlattices \cite{Serra2013}. In such systems, disclination lines loop around the solid in many possible linkages. A field above threshold rewires them into a new linkage that persists after the field is removed, storing memory in the topology of the defects \cite{Araki2011}. Cubic microlattices designed to demonstrate this effect retain field-written states for days \cite{Serra2013}.
Switching fields can drive such systems into a metastable state; melting returns them to an isotropic state, and quenching can define a new state.
In particularly interesting circumstances, nematic structures can be switched from one state to another and back again, e.g. by rewiring disclinations with in-plane fields \cite{Harkai2020}, by flexoelectric pulses at a grating surface \cite{BryanBrownGABN,Spencer2010} or by shear flow over micropillars \cite{Kara2025}. Given these capabilities, reconfigurable defect systems have been employed as bistable low-power displays \cite{BryanBrownGABN,Spencer2010}, tunable diffraction gratings \cite{KimSerra2020} and as reconfigurable memory \cite{Serra2013,Harkai2020,Kara2025}.

The ability to manipulate defects as we do colloids would make them addressable building blocks for such devices.  For example, colloids can dock near curved walls with similar anchoring via lock-and-key interactions, guided by splay and bend distortions in the domain \cite{Silvestre2004,EskandariPSTT,Silvestre2014,Luo2016,SilvestreTasinkevych2017,Luo2018,Edziah2026}. In this interaction, the colloid is a foreign inclusion with a distinct topological charge which distorts the nematic through its curvature and its anchoring energy. While defects can also be stabilized near curved walls, it is not obvious a priori that defects should participate in lock-and-key interactions. Unlike colloids, topological defects are intrinsic to the nematic order, with a nanometric core in which order is lost, surrounded by strong director distortion \cite{Kleman2006}. While anchoring dictates the nematic director field around a colloid, the director field around a defect core results from energy minimization. Furthermore, bulk defects can transform from one form to another; a radial and a hyperbolic hedgehog, for example, are related by a continuous rotation of the director \cite{Alexander2012}. Confinement with strong anchoring penalizes such director rotation. Radial and hyperbolic hedgehogs then become distinct states, separated by the elastic energy needed to reorganize the texture against the anchoring. In a given setting, the number, charge, structure and stability of the defects that form depend on topological constraints and on tradeoffs among elasticity, surface anchoring and the shape of the confining vessel \cite{Kleman2006,Kralj1995,Tran2016}.

Here, we introduce topological bits, defects stabilized in pillbox-shaped pores with homeotropic anchoring. Such pore geometries have been used to develop multistable elastic pixels, in which a colloid is toggled between stable lock-and-key states by switching fields \cite{Edziah2026}. Absent colloids, under strong homeotropic anchoring, a defect can form in each pore. The defect, a radial hedgehog, sits in one of two curved lobes, in lock-and-key energy wells. A DC electric field along the pore axis toward the opposite lobe moves the defect into that lobe, where it stays once the field is off (Figure~\ref{fig:overview}). Migration is linear in the field and reverses with polarity, consistent with flexoelectric coupling. Within a flexoelectric description, the nematic surrounding the radial defect carries a net positive bulk bound charge which is balanced by negative bound charges near the pore walls. Their spatial separation produces a net electric dipole, which equivalently derives from integrating the flexoelectric polarization density over the pore domain. An electric field switches the defect position by increasing the alignment of the net dipole with the applied field. Pores with different waist curvature switch at different thresholds, allowing arrays of topological bits to encode binary information. Manipulating defects as charged objects offers a route to field-addressable, reconfigurable soft materials.
\section{Results and Discussion}

\subsection{Pillbox pores}

\begin{figure}
\centering
\includegraphics[width=\textwidth]{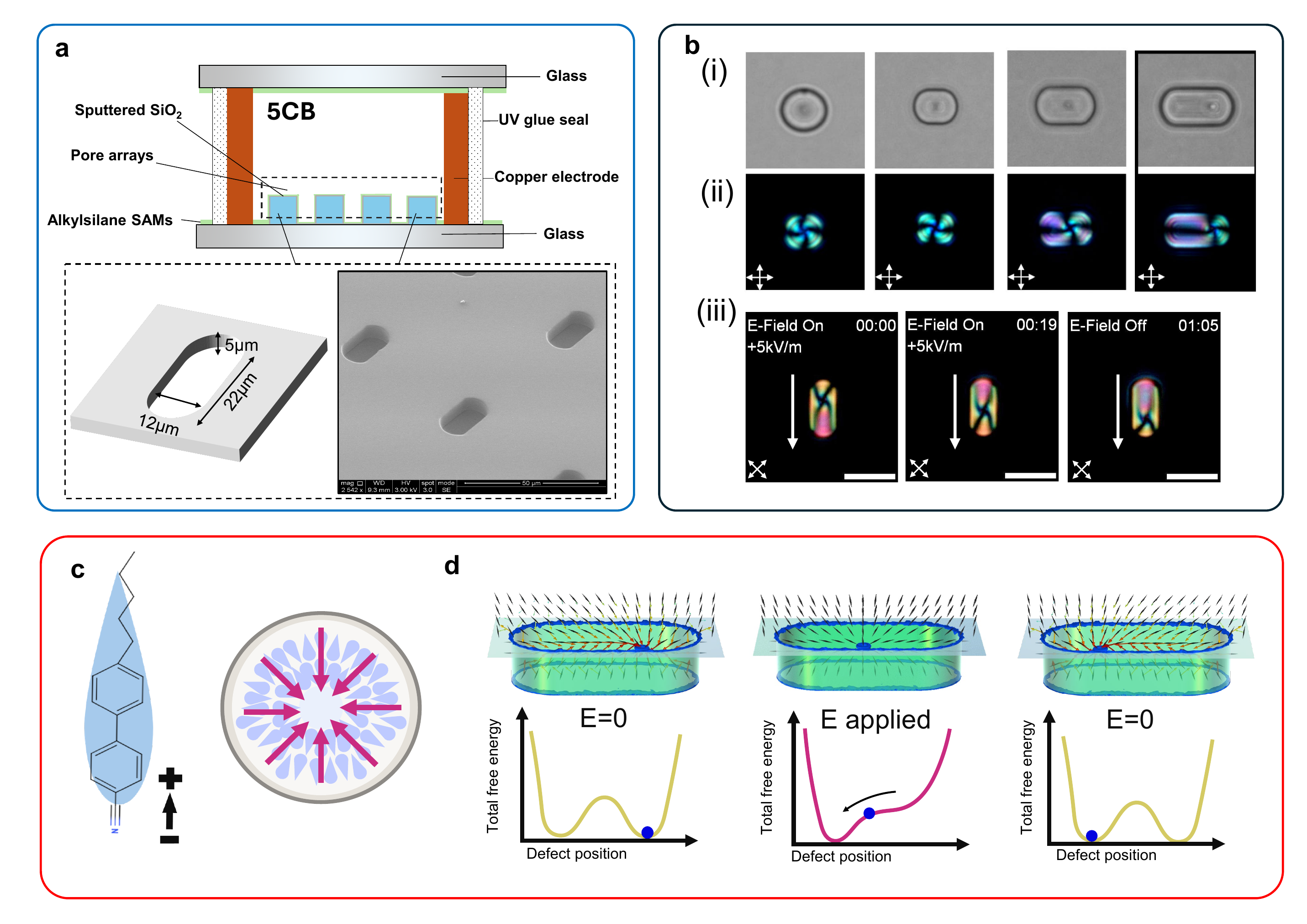}
\caption{Field-driven reconfiguration of point defects confined in pillbox pores. Point defects in SU-8 pores translate along an applied DC field and remain trapped in their new positions after the field is removed. (a) Sandwich cell with \SI{50}{\micro\meter} copper strips serving as spacers and in-plane electrodes, \SI{1}{\centi\meter} apart. All surfaces are treated with alkylsilane surfactants for homeotropic anchoring. Schematic of a single pore (depth \SI{5}{\micro\meter}) and scanning electron micrograph of the pore array. (b) Bright-field images (i) and crossed-polarizer textures (ii) of defects in pores of increasing elongation with fixed end radii. (iii) Time-resolved POM images of field-driven reconfiguration at \SI{5}{\kilo\volt\per\meter} with polarizers placed at $\mathrm{45}^{\circ}$ with respect to the sample. The field was turned off at $t = \SI{64}{\second}$. Inserted scale bars are \SI{20}{\micro\meter}.(c) 5CB carries a permanent dipole along its long axis. In the splayed director field of a radial hedgehog, the dipoles orient toward the core, giving a flexoelectric polarization directed toward the defect. (d) Landau-de Gennes simulations of the director field and schematic free-energy profiles before, during and after switching, showing the lock-and-key wells and their tilt by the field. }
\label{fig:overview}
\end{figure}

The pores have a pillbox cross section, with curved ends connected by straight sidewalls, patterned in SU-8 on glass (Figure~\ref{fig:overview}a). Pores are $h=\SI{5}{\micro\meter}$ deep, $a = \SI{22}{\micro\meter}$ long, $b=\SI{12}{\micro\meter}$ wide, and the lobes have a radius of curvature of \SI{9}{\micro\meter}. The pore walls, floor and the upper glass plate are functionalized with DMOAP to impose homeotropic anchoring. Copper strips \SI{50}{\micro\meter} thick serve as both spacers and in-plane electrodes, so that a layer of 5CB separates the pore array from the upper plate (Figure~\ref{fig:overview}a). We define an $x$ coordinate along the pore axis, with origin at the pore midpoint.

In circular pores, three configurations are accessible, an escaped texture without singularity and textures containing a point defect (Supporting Information, Figure~S1). Bright-field and crossed-polarizer images distinguish escaped from defect-bearing textures, but cannot distinguish a radial from a hyperbolic hedgehog, since their projections are identical. Under a strengthened homeotropic treatment, in which surfaces are activated by \ce{O2} plasma before silanization, point defects form in all pores and escaped textures are absent.

\subsection{Defect lock-and-key}

\begin{figure}
\centering
\includegraphics[width=0.85\textwidth]{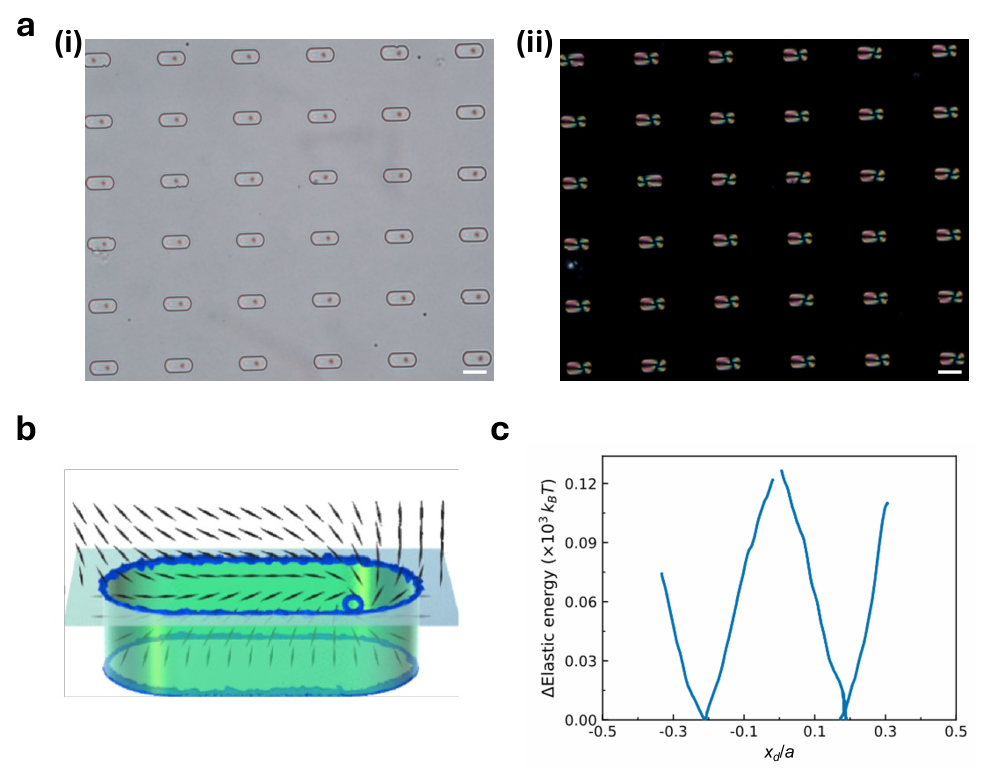}
\caption{Lock-and-key wells. (a) Bright field (i) and the corresponding POM (ii) images of the radial hedgehog defect in the pillbox arrays. The cross polarizer is placed at $\mathrm{90}^{\circ}$ with respect to the sample. (b) Landau-de Gennes director field of a defect docked in a lobe. (c) Zero-field elastic energy profile reconstructed from defect relaxation trajectories, versus defect position $x_d$ normalized by the pillbox length $a$. Two lobe minima are separated by a barrier at the pore center. The inserted scale bars are 20$\mathrm{\mu}m$.}
\label{fig:lockkey}
\end{figure}

Elongating a circular pore into a pillbox breaks its symmetry and creates two docking sites, one in each lobe (Figure~\ref{fig:overview}b). As the aspect ratio $a/b$ increases from 1 to 1.83, 
 point defects increasingly localize in either lobe. Once localized, a defect remains in its lobe without spontaneous migration over the period of observation. Landau-de Gennes simulations using 5CB elastic constants and a scaled-down pore geometry that preserves the experimental aspect ratios reproduce this localization of the radial hedgehog (Figure~\ref{fig:lockkey}b). In the simulations, the pore has length $a = \SI{1}{\micro\meter}$, width $b = \SI{500}{\nano\meter}$ and depth $h = \SI{250}{\nano\meter}$, so that $a/b = 2$ and $a/h = 4$. These aspect ratios are similar to those of the experimental pores. The lobes have radius of curvature $R = b/2$. Thus, while the ratio $R/b$ is similar to that in experiment, the absolute lobe curvature, $1/R$, is greater in the simulations.

The lock-and-key interaction here is related to that of colloids \cite{Luo2016,Luo2018}. While no particle is present, the distortion from the splay field of the radial hedgehog (``the key'') is minimized when the defect sits adjacent to the curved lobe (``the lock''), reducing the elastic free energy. Relaxation trajectories give an energy profile with two lobe minima separated by a barrier of about $120\,k_BT$ at the pore center (Figure~\ref{fig:lockkey}c) for a $C_D^* = 1\ \mu\mathrm{N\ s/m}$.

\subsection{The lobe defect is a radial hedgehog}

\begin{figure}
\centering
\includegraphics[width=\textwidth]{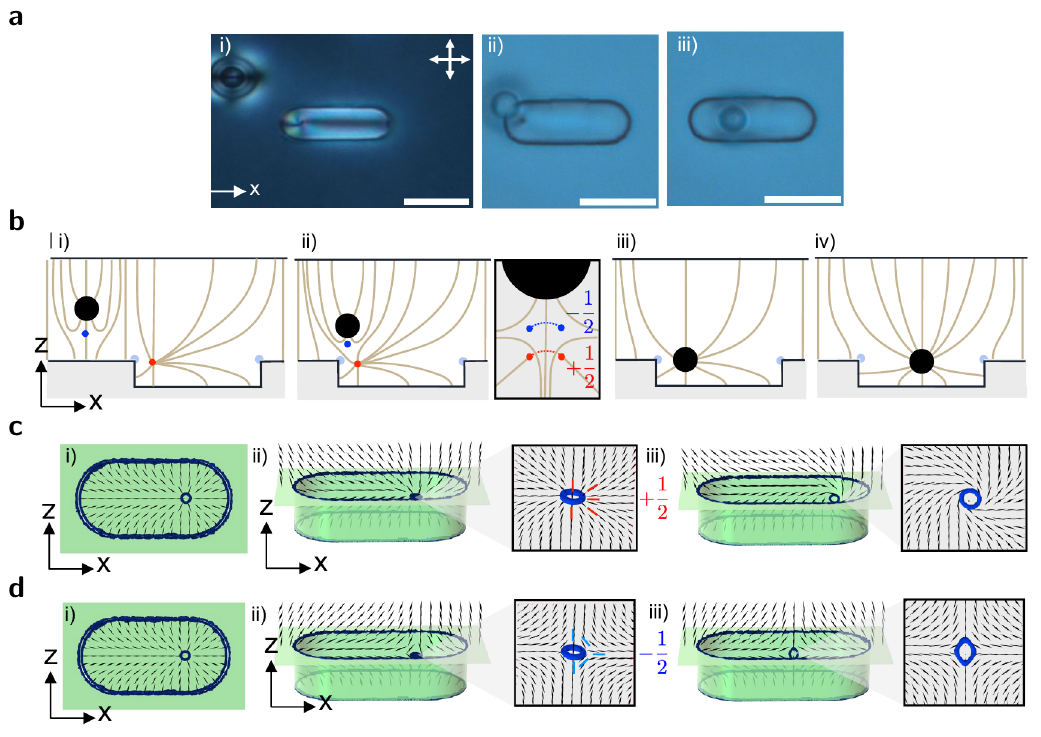}
\caption{Identification of the lobe defect. (a) A homeotropic colloid assembling with the lobe defect. (i) Crossed polarizers, colloid with its companion hyperbolic hedgehog approaching the pore. (ii,iii) Bright field, the colloid enters the lobe and the companion defect annihilates with the lobe defect. Scale bars, \SI{20}{\micro\meter}. (b) Side-view schematic of the director during assembly. Blue, companion hyperbolic hedgehog ($-1/2$ ring). Red, lobe radial hedgehog ($+1/2$ ring). Pale blue, $-1/4$ edge disclinations at the pore rim. (c,d) Landau-de Gennes simulations of a radial hedgehog (c) and a hyperbolic hedgehog (d). (i) Top view of the director in the pore midplane, one-constant approximation. (ii) Side view, one-constant approximation. (iii) Side view, three-constant approximation. Insets show the defect core, a small disclination ring of winding $+1/2$ (radial) or $-1/2$ (hyperbolic). With three constants, the radial hedgehog remains in the lobe, while the hyperbolic hedgehog retreats toward the pore center.}
\label{fig:identity}
\end{figure}

We identified the experimental defect that forms under strong anchoring and localizes in the lobe by using a homeotropic silver-coated silica colloid (\SI{9}{\micro\meter} in diameter) as a topological probe in the absence of applied electric field  (Figure~\ref{fig:identity}a,b). The colloid carries net topological charge of $+1$, balanced by a companion hyperbolic hedgehog in a dipolar configuration. As the colloid enters the lobe, the companion hyperbolic hedgehog annihilates with the lobe defect in an isolated event, leaving the pore-edge disclinations and the surrounding director undisturbed.  A hyperbolic lobe defect would carry the same charge as the companion and could not annihilate without a global rearrangement. The lobe defect is therefore a radial hedgehog. Landau-de Gennes simulations support this identification. We minimized the free energy in a domain representing the pore, initialized with either a radial or a hyperbolic hedgehog (Figure~\ref{fig:identity}c,d). In both cases the relaxed core is a small disclination ring rather than a point singularity, a $+1/2$ wedge ring for the radial hedgehog and a $-1/2$ ring for the hyperbolic one.

Lobe localization in simulation requires weak anchoring at the upper boundary of the domain. This free boundary stands in for the deep nematic layer between the pore and the distant upper plate. Thus, the simulations reveal that defect localization is sensitive to the upper-boundary anchoring. In simulations with one elastic constant, both radial and hyperbolic hedgehogs relax into a lobe (Figure~\ref{fig:identity}c(i,ii) and d(i,ii), respectively). With three distinct elastic constants for 5CB, the radial hedgehog remains in the lobe, while the hyperbolic hedgehog retreats toward the pore center (Figure~\ref{fig:identity}c(iii) and d(iii), respectively). 

\subsection{Field-driven migration between lobes}

\begin{figure}[H]
\centering
\includegraphics[width=1\textwidth]{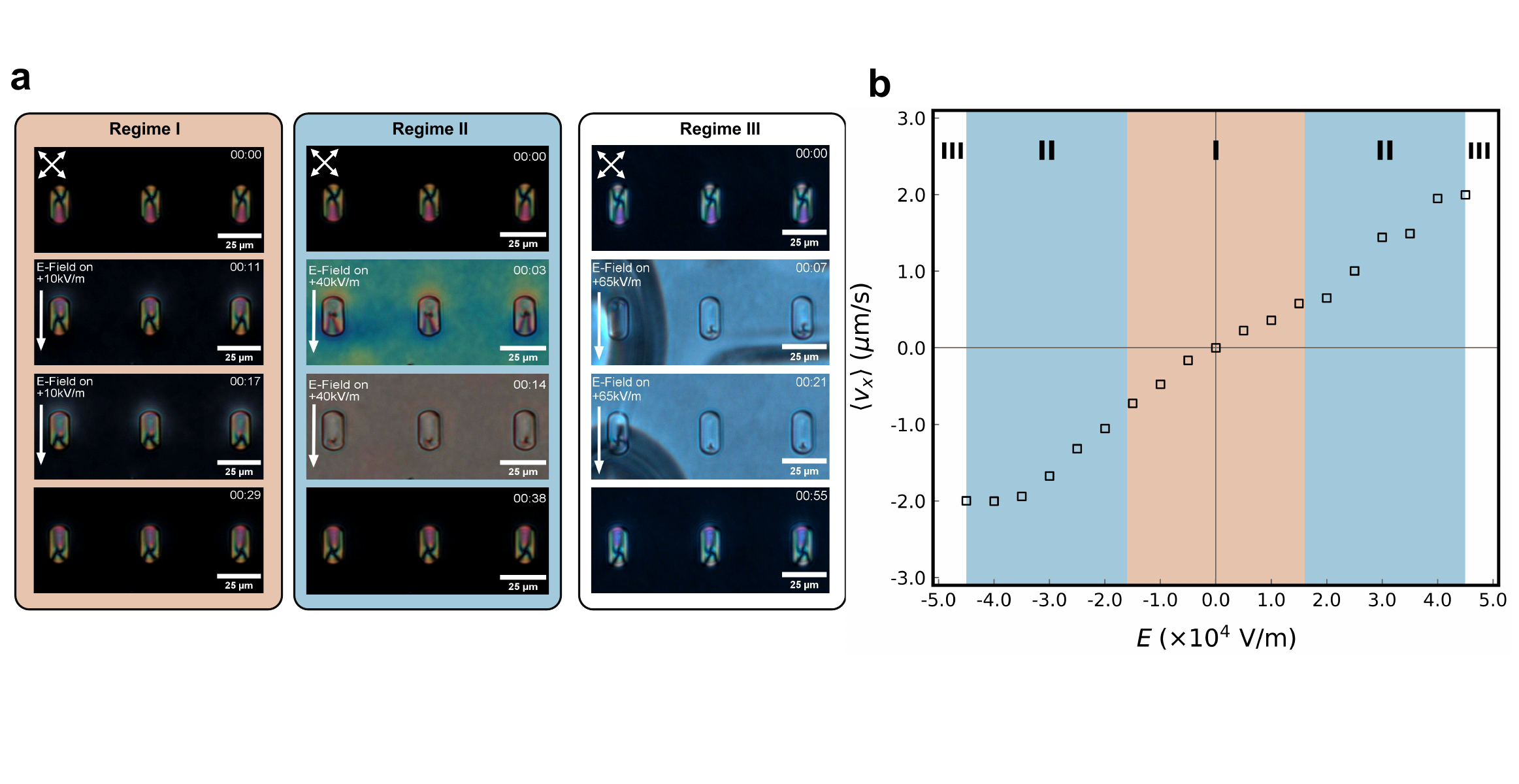}
\vspace{-60pt}
\caption{Field-dependent regimes of migration of confined radial hedgehogs. (a) Crossed-polarizer images in regimes I, II and III, sample at \SI{45}{\degree} to the polarizer. (b) Mean migration velocity $\langle v_x\rangle$ versus applied field. Migration reverses with field polarity. Regime I, $|E_x| \lesssim \SI{15}{\kilo\volt\per\meter}$, reproducible switching with reorientation localized in the pore. Regime II, reorientation above the pores visible, synchronized switching maintained. Regime III, domain-wall formation and decay with a nonlinear velocity response.}
\label{fig:regimes}
\end{figure}

A DC field applied along the pore axis and directed toward the opposite lobe pulls the defect out of its initial lobe and drives it into the other (Figure~\ref{fig:overview}b(iii)). The defect moves in the direction of the field, and reversing polarity reverses its motion. Without a field it stays docked. Switching is smooth and reproducible for $|E_x| \le \SI{45}{\kilo\volt\per\meter}$, and over this range the mean velocity increases approximately linearly with field strength (Figure~\ref{fig:regimes}b).

The response falls into three regimes. In regime I ($|E_x| \lesssim \SI{15}{\kilo\volt\per\meter}$), director reorientation is confined primarily to the pore. In regime II, the nematic in the homeotropic gap above the pores undergoes a Fréedericksz transition, producing director reorientation visible under crossed polarizers, while defect switching remains synchronized across the array. In regime III ($|E_x| \gtrsim \SI{45}{\kilo\volt\per\meter}$), domain walls form between regions with oppositely tilted splay–bend director configurations generated by the Fréedericksz transition in the homeotropic gap. These walls interact with the confined defects as they evolve and annihilate, producing a nonlinear velocity response. If the field is maintained until the walls have annihilated, the confined defects again move to the lobe selected by the field. 

Dielectric coupling and the quadratic component of liquid-crystal-enabled electrophoresis \cite{Lavrentovich2010,Lazo2013} are even in the field and cannot reverse the direction of migration with polarity, so they cannot account for the observed motion. A defect would move linearly with the field by electrophoresis if it contained ions, which can, in principle, partition into defect cores \cite{Everts2021}. Flexoelectric polarization, by contrast, requires no free charge. It arises wherever the director splays or bends, throughout the confined nematic, and it couples linearly to the field. We show that the space charge distribution associated with flexoelectricity can account for the observed migration.

\subsection{Defect trajectories and the energy landscape}

To interpret defect migration, we use an effective description based on the defect position $x_d(t)$ and the total free energy of the NLC in presence of the electric field $U(x_d)$. The zero-field elastic energy $U_\mathrm{elas}(x_d)$ forms a double-well landscape, with a minimum in each lobe and a barrier at the pore center. The elastic landscape was inferred from the relaxation of a defect released near this barrier toward a lobe (Fig. \ref{fig:lockkey} c). Here, we apply the same approach to field-driven trajectories to reconstruct the total effective energy $U(x_d)$ and separate its elastic $U_\mathrm{elas}(x_d)$ and field-induced contributions $U_{\mathrm{field}}(x_d)$.

\begin{figure}
\centering
\includegraphics[width=\textwidth]{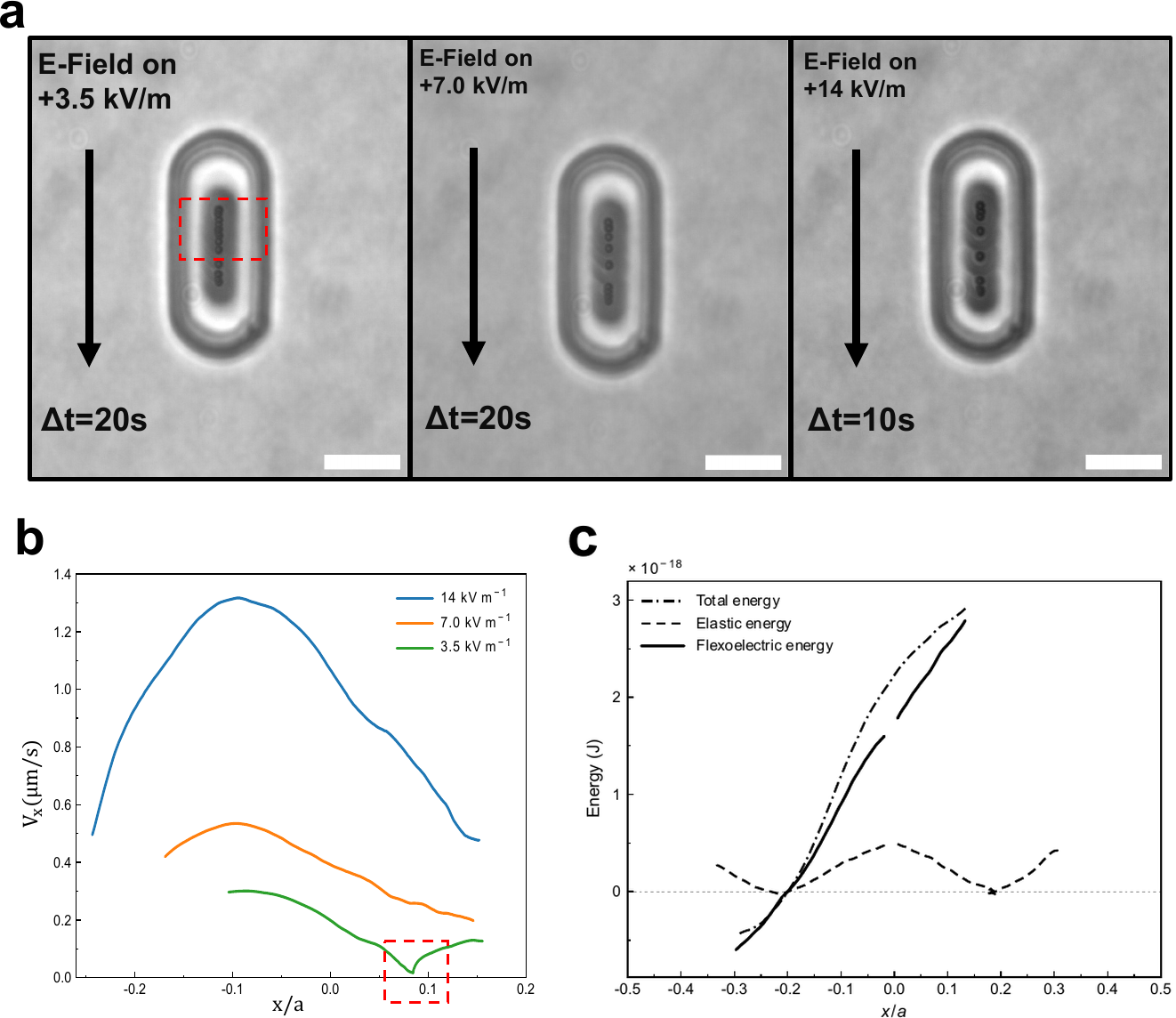}
\caption{Migration at low field and reconstruction of the energy landscape. (a) Superimposed bright-field images of migration at $E_x = 3.5$, 7 and \SI{14}{\kilo\volt\per\meter}, with intervals of 20, 20 and \SI{10}{\second} between images. Arrows indicate the field direction. The red box marks the near-stall at \SI{3.5}{\kilo\volt\per\meter}. (b) Instantaneous velocity versus $x_d/a$, with $x_d$ measured from the pore center and $a$ the pillbox length. (c) Energy landscapes reconstructed at $E_x = +\SI{5}{\kilo\volt\per\meter}$. Total (dash-dot), elastic (dashed) and flexoelectric (solid) contributions. The flexoelectric contribution is the difference between field-on and zero-field landscapes. Scale bars, \SI{10}{\micro\meter}.}
\label{fig:landscape}
\end{figure}

We tracked migration at \SI{20}{fps} at $E_x = 3.5$, 7 and \SI{14}{\kilo\volt\per\meter}, within regime I (Figure~\ref{fig:landscape}a,b). At \SI{3.5}{\kilo\volt\per\meter}, the defect slows significantly near the pore center and oscillates transversely for about \SI{20}{\second} before crossing, after which its speed increases toward the receiving lobe. At 7 and \SI{14}{\kilo\volt\per\meter}, no such stalling is observed. Rather, the velocity peaks near $x_d/a = -0.1$ and decreases as the defect enters the receiving lobe. At all applied field strengths, the defect leaves the original lobe more slowly than it approaches the far one, consistent with the elastic force opposing migration away from the old docking site and aiding migration toward the new one.
Neglecting inertia, the change in energy along a trajectory follows from the velocity,
\begin{equation}
\frac{\Delta U(x_d)}{C_D^*} = -\int_{x_{d,\mathrm{ref}}}^{x_d} v_x(x_d')\,dx_d',
\label{eq:landscape}
\end{equation}
where $C_D^*$ is an effective drag coefficient, taken to be the same with and without the field. Subtracting the zero-field elastic landscape from the field-on landscape at \SI{5}{\kilo\volt\per\meter} isolates the electric field contribution $\Delta U_\mathrm{field}(x_d)$(Figure~\ref{fig:landscape}c), where $C_D^* \approx \SI{1}{\micro\newton\second\per\meter}$. This contribution is linear in defect position, so the field exerts a constant force on the defect. In regime I, the field strength is below the Fr\'eedericksz transition in the gap. The electric field tilts the total energy $U(x_d)$, causing defect migration toward the receiving lobe, the unique minimum. When the field is switched off, the zero-field double-well landscape is restored: a defect that has entered the receiving lobe relaxes into its elastic minimum and remains trapped there.

\subsection{Flexoelectric defect migration}

Flexoelectric polarization arises from splay and bend of the director, described by the splay vector $\mathbf{s} = \mathbf{n}(\nabla\cdot\mathbf{n})$ and the bend vector $\mathbf{b} = (\mathbf{n}\cdot\nabla)\mathbf{n}$ \cite{Meyer1969},
\begin{equation}
\mathbf{P} = e_1\mathbf{s} + e_3\mathbf{b}.
\label{eq:meyer}
\end{equation}
The polarization produces a bound charge density $\rho_b = -\nabla\cdot\mathbf{P}$ in the bulk and $\sigma_b = \mathbf{P}\cdot\bm{\nu}$ on a boundary with outward normal $\bm{\nu}$.

To motivate our discussion, consider a radial hedgehog, a defect with pure splay. For $e_1 < 0$ its polarization points toward the core (Figure~\ref{fig:overview}c), and $\rho_b = -2e_1/r^2$ is positive throughout the surrounding nematic. A sphere of radius $R$ wrapping the defect would enclose the charge $\sim -8\pi e_1 R$. The nematic near the defect therefore carries a positive bound charge, compensated by negative bound charge in the domain and on its boundaries. While this picture motivates the concept of positive bound charge in regions of strong splay, assigning a charge to the defect in this way raises two difficulties. First, the enclosed charge grows with $R$, so it depends on an arbitrary cutoff radius. Second, the ideal radial hedgehog is a simplified view of a $+1$ defect. In Landau-de Gennes simulations, the core of a $+1$ defect relaxes into a small disclination ring, with a more complex director field. 

\begin{figure}
\centering
\includegraphics[width=\textwidth]{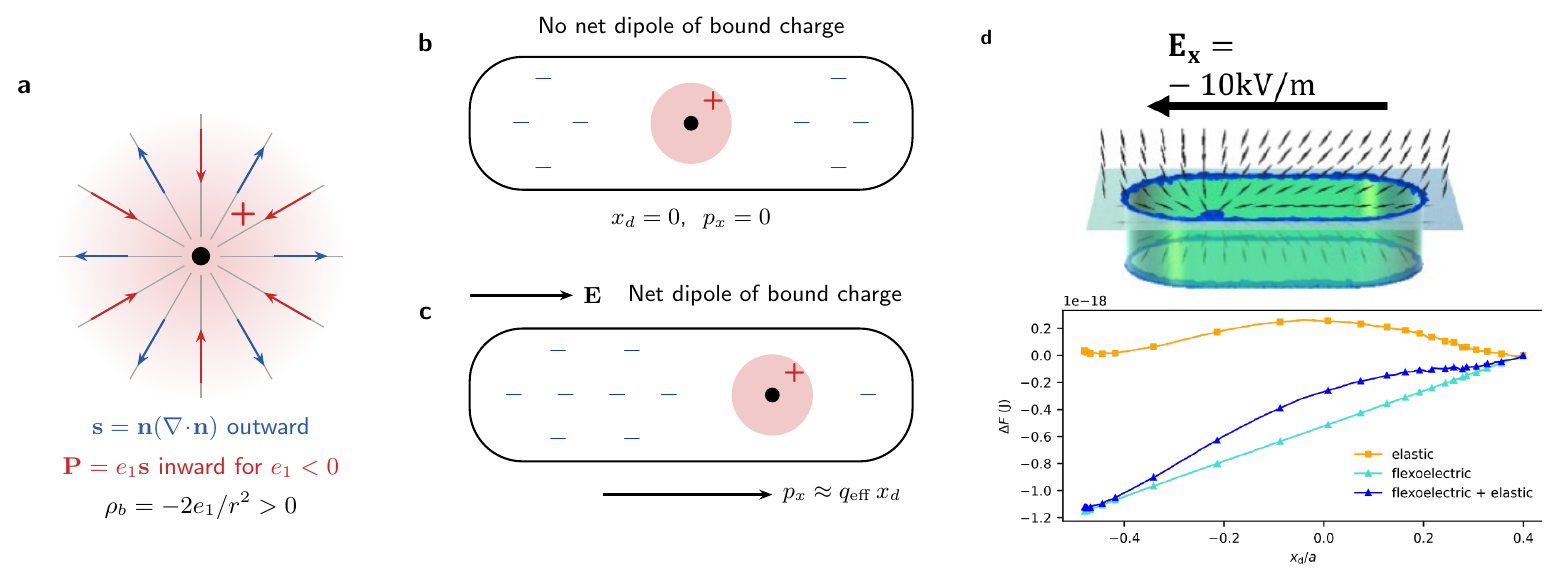}
\caption{Flexoelectric mechanism of defect migration. (a) Radial hedgehog. The splay vector $\mathbf{s}$ points outward from the core (blue). For $e_1 < 0$ the splay polarization points inward (red), and the bound charge density $\rho_b$ is positive in the surrounding nematic. (b) Defect at the pore midpoint. Its positive partial charge is compensated symmetrically by negative bound charge in the rest of the pore and on its walls, and the net dipole vanishes. (c) Defect displaced along $\mathbf{E}$. The bound charge acquires a first moment, a dipole $p_x$ that grows with displacement. (d) Landau-de Gennes simulation of switching at $E_x = -\SI{10}{\kilo\volt\per\meter}$. Elastic, flexoelectric and total free energy versus defect position $x_d/a$, with the simulated director field in the pore. The simulation illustrates the mechanism and is not a scale model of the experiment.}
\label{fig:mechanism}
\end{figure}

To avoid both difficulties, we consider instead the net dipole of the pore,
\begin{equation}
\mathbf{p}(x_d) = \int_V \mathbf{P}\,dV,
\end{equation}
which requires neither a cutoff nor a simplified picture of the defect. The net dipole is a state function of the defect position $x_d$. For field strengths well below the Fr\'eedericksz threshold, where dielectric contributions are negligible, the free energy of the nematic can be expressed as the sum of elastic and flexoelectric contributions,
\begin{equation}
U(x_d) = U_\mathrm{elas}(x_d) - E_x\,p_x(x_d),
\end{equation}
where $E_x$ is the $x$ component of the electric field $\mathbf{E}$. This allows us to define the associated flexoelectric force as $F_\mathrm{flex} = q_\mathrm{eff}E_x$, with
\begin{equation}
q_\mathrm{eff} = \frac{dp_x}{dx_d}
\label{eq:qeff}
\end{equation}
being the effective charge of the defect. The defect moves in the field to increase the net dipole in the pore. For a splay-rich structure, with the defect at the midpoint, its positive partial charge is compensated symmetrically and $p_x = 0$. Displacing it gives the bound charge a first moment, a dipole that grows as $q_\mathrm{eff}x_d$, as for a mobile positive charge (Figure~\ref{fig:mechanism}). The predicted force is linear in $E_x$ and is defined by the nematic texture in the pore. 

In experiment, for the assumed  $C_D^*$, the slope of the flexoelectric energy in Figure~\ref{fig:landscape}c gives a force of about \SI{4e-13}{\newton} at \SI{5}{\kilo\volt\per\meter}, so $q_\mathrm{eff} \approx \SI{8e-17}{\coulomb}$, about 500 elementary charges. Simulations performed in the Landau-de Gennes framework, including elastic and flexoelectric effects, for 5CB with $G_i = \SI{-10}{\pico\coulomb\per\meter}$ at $E_x = \SI{-10}{\kilo\volt\per\meter}$ demonstrate flexoelectrically driven defect motion. The simulated pore is not scaled to the experimental one (Methods), since resolving the nanometric defect core over tens of micrometers is computationally prohibitive, and its upper boundary is free, standing in for the deep nematic layer above the pore. In addition to splay and bend contributions, simulations also address contributions of gradients in order to the flexoelectric effect (Methods). Despite differences in scale and lid anchoring, the simulations show essentially the same phenomenon. Like experiment, the flexoelectric energy is linear in defect position, tilts the total energy landscape, and drives the defect to the lobe selected by the field (Figures~\ref{fig:overview}d and \ref{fig:mechanism}d).

As $E_x$ increases, dielectric effects become important in the gap. Above the Fr\'eedericksz threshold, $E_F = (\pi/d)\sqrt{K_3/(\varepsilon_0\Delta\varepsilon)} \approx \SI{20}{\kilo\volt\per\meter}$ for the gap of thickness $d \approx \SI{50}{\micro\meter}$ \cite{Bogi2001}, the nematic in the gap reorients to align with the applied electric field, marking the onset of regime II. In this regime, the nematic texture in the pore is protected from dielectric reorientation by the strong anchoring on its base and walls.

Migration of defects in DC fields has been reported previously for dowsons, special defects which form in homeotropic thin films. Dowsons are point defects of the dowser texture, a quasi-planar nematic state characterized by a vector at its midplane; they move in electric fields through the coupling of this vector to the field \cite{Pieranski2019,Pieranski2020}. In our system, the splay and bend of the radial hedgehog itself couple to the field. It would be interesting to extend this work to other defect structures, which have distinct splay and bend signatures and couple differently to the external field. For example, unlike the ideal radial hedgehog, the hyperbolic hedgehog has a weak monopolar bound charge, with dominant quadrupolar and higher-order moments. Since a uniform field couples only to the monopolar part, a highly attenuated flexoelectric response is predicted. In our simulations, the hyperbolic hedgehog responds only weakly to $E_x$, in keeping with the predicted attenuated response. In our framework, since splay and bend around the hyperbolic hedgehog vary, its effective charge, defined as the change in the net dipole of the domain with defect position, is a tensor, and its displacement need not follow the field. Related translations have been predicted for Dowsons \cite{Pieranski2020}. 


\subsection{Geometry-programmed thresholds and binary encoding}

\begin{figure}
\centering
\includegraphics[width=\textwidth]{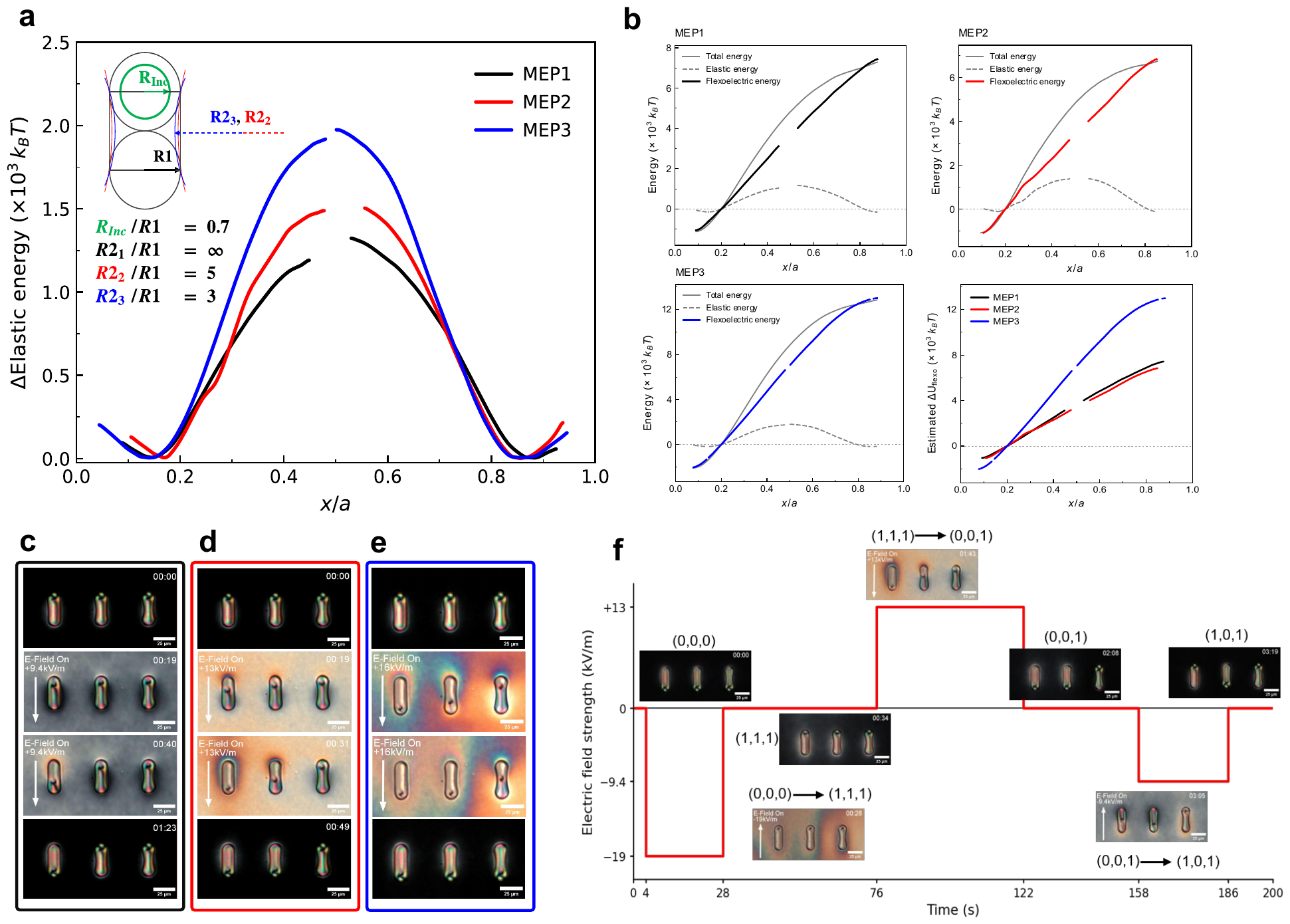}
\caption{Geometry-dependent energy landscapes and selective switching of binary states. (a) Zero-field elastic energy landscapes for pores 1 (black), 2 (red) and 3 (blue). Inset, geometry with $R_\mathrm{inc}/R_1 = 0.7$ and $R_{2}/R_1 = \infty$, 5 and 3. (b) Total (gray solid), elastic (gray dashed) and estimated flexoelectric (colored) energies for each geometry, and an overlay of the flexoelectric estimates. Defect positions $x_d$  are normalized by $a \approx \SI{42}{\micro\meter}$. Energies with $C_D^* = \SI{1}{\micro\newton\second\per\meter}$.  (c--e) Switching under $+9.4$, $+13$ and $+\SI{16}{\kilo\volt\per\meter}$, of pore 1, pores 1 and 2, and all three. Arrows indicate the field direction. (f) Field sequence and binary states. States 0 and 1 denote occupancy of the lower and upper lobes. Scale bars, \SI{25}{\micro\meter}.}
\label{fig:encoding}
\end{figure}

Pore shape sets the barrier between lobes. We kept the lobe curvature fixed and added antagonistic curvature at the pore waist, with $R_2/R_1 = \infty$, 5 and 3 for pores 1, 2 and 3 (Figure~\ref{fig:encoding}a). These pores are longer than the pore studied in Figures~\ref{fig:lockkey} and \ref{fig:landscape}, with $a \approx \SI{42}{\micro\meter}$. The waist curvature adds splay and bend at the pore center that accentuates the incompatibility with the director around a radial hedgehog defect, raising the elastic energy barrier by about $800\,k_BT$ for pore 3 compared with pore 1. As a result, curvature at the waist steepens the elastic energy profile between the minima and the midpoint, increasing the maximum restoring elastic force. In all three pores, the flexoelectric energy remains linear in defect position, corresponding to an approximately constant driving force at a given field. In the deterministic, overdamped description, switching requires this driving force to exceed the maximum opposing elastic force (Figure~\ref{fig:encoding}b). Thus, the threshold electric field strength to achieve switching increases with the curvature of the pillbox waist. Each pore switches at its own threshold field, \SI{9.4}{\kilo\volt\per\meter} for pore 1, \SI{13}{\kilo\volt\per\meter} for pore 2 and \SI{16}{\kilo\volt\per\meter} for pore 3, for either polarity (Figure~\ref{fig:encoding}c--e). 

 Each pore encodes one bit of information, with state 1 corresponding to a defect in the upper lobe and state 0 to a defect in the lower lobe. We refer to this unit as a topological bit because its information carrier is a topological defect; the two logical states are topologically equivalent, and information is encoded in defect position. The three pores therefore form a three-bit array. Successive square electric-field pulses of $-19$, $+13$ and $\SI{-9.4}{\kilo\volt\per\meter}$ drive the array from $(0,0,0)$ to $(1,1,1)$, then $(0,0,1)$ and finally $(1,0,1)$, with each configuration retained after the field is removed (Figure~\ref{fig:encoding}f). The distinct switching thresholds enable selective writing through the sign and amplitude of a common applied field.

These results expand a family of defect-based memory and computation schemes. Generically, defect structures encode information in their topology and configuration. The encoding of Campos-Villalobos et al. is a form of read-only memory, with the 0s and 1s built into the fabricated structure, although dynamic rewriting is predicted through control of anchoring \cite{CamposVillalobos2026}. The nematic bits of Kos and Dunkel are continuous, qubit-like states \cite{KosDunkel2022}, bits in a broader sense than a binary digit. Here, we demonstrate topological bits, which act as binary digits, similar to magnetic bits. There are two states, the defect in one lobe or the other, written by the sign of the field, retained without power, and read optically.

\section{Conclusions}

We stabilize a radial hedgehog in a pillbox pore with curved lobes at either end. The defect docks in one lobe or the other, in lock-and-key energy wells set by the curved walls. Under a DC external field, the defect moves from one lobe to the other with average speed linear in the field, and remains there once the field is removed. We discuss this result in the context of the flexoelectric effect. The nematic near the defect carries a positive flexoelectric bound charge, compensated in the rest of the pore and on its walls. The net dipole of the liquid crystal in the pore depends on the position of the defect. The DC field drives the defect in the direction that increases that dipole. Antagonistic curvature at the pore waist raises the switching threshold, so pores of different shape switch at different fields. Each pore thus serves as a “topological bit,” with its binary state encoded in which lobe hosts the topological defect. Arrays of topological bits can be written and rewritten with field pulses.

These topological bits have several strengths. Defect-filled arrays form spontaneously and deterministically, encoding nonvolatile information by localizing the radial hedgehog in one of the two lobes of each pore. Although the experimental lid configuration does not topologically require a defect in the pore, the nematic forms the defect texture with high probability under the current anchoring treatment and quench protocol. This robust outcome does not imply that the defect state is the ground state of the energy landscape, but suggests that the formation pathway strongly favors relaxation into its basin of attraction. Once formed, the defect persists through the distortions associated with switching between lobes. Energy barriers then impede relaxation to alternative topologies. In closed pores, where defects are required, the defect enjoys topological protection. In either case, the two logical states are topologically equivalent: retention of the encoded information relies on the elastic barrier between the two defect positions. Pore shape sets the switching threshold, so a single uniform field can address subsets of pores selectively without patterned electrodes. Since the velocity is linear in field strength, the topological bits' stable states can be switched by changing the field polarity. Switching occurs below the Fr\'eedericksz threshold of the surrounding nematic, and every pore can be read optically in parallel. Simulation at the scale of hundreds of nanometers suggests that the system can be scaled down and preserve these features. The defect is predicted to move as an effective charge set by the flexoelectric dipole of the pore, with no free ions required. While we focused our experiments on the strong anchoring regime, weaker anchoring may allow more complex encoding in a single array, with escaped textures as fixed bits, hyperbolic hedgehogs as complex, weakly responsive bits, and radial hedgehogs moving from one non-volatile site to another in external fields.

Several directions remain open. First, molecular dynamics simulations are underway to establish the molecular origin of flexoelectric polarization in 5CB. In Meyer's picture, distortion polarizes the nematic by biasing the direction of permanent dipole moments on nematogens \cite{Meyer1969}. However, 5CB carries a strong permanent dipole and a propensity to form antiparallel dimers. These may need to be disrupted in regions of splay or bend to impart flexoelectric coupling. Second, free ions could play important roles. They would generate strong electro-osmotic and electrophoretic motions, and could also screen the flexoelectric bound charge. Ions can partition into defect cores \cite{Everts2021}, where their charge would compete with the bound charge of the defect. Defects collect other cargo as well, including colloids and solutes; whether flexoelectric migration can shuttle such cargo between sites is an interesting open issue. Third, while we study the radial hedgehog with its strong monopolar bound charge, the hyperbolic hedgehog, with its predominantly quadrupolar charge distribution, has interesting open degrees of freedom in its tensorial effective charge and resulting flexoelectric response. 

\section{Methods}

\subsection{Substrate preparation}
Photomasks defining circular and elongated pores were designed in LayoutEditor and written by direct-write laser lithography at the Singh Center for Nanotechnology. Pore arrays were patterned on \SI{4}{in.} glass wafers in the negative epoxy photoresist SU-8 3005 (Kayaku Advanced Materials), with spin-coating, baking and exposure following the manufacturer's guidelines. Because DMOAP adsorbs poorly to SU-8, cured substrates were coated with \SI{20}{\nano\meter} of \ce{SiO2} in a Kurt J. Lesker PVD 75 DC/RF sputtering system. Upper glass substrates were cleaned in piranha solution for \SI{1}{\hour}, rinsed with deionized water and dried. SU-8 and glass substrates were etched in \ce{O2} plasma for \SI{60}{\second}, immersed in a \SI{3}{wt\percent} aqueous solution of dimethyloctadecyl[3-(trimethoxysilyl)propyl]ammonium chloride (DMOAP, Gelest) for \SI{10}{\minute} at room temperature, rinsed with deionized water and baked at \SI{80}{\celsius} for \SI{30}{\minute}. In earlier experiments, in which escaped and defect-bearing textures coexisted (Supporting Information), the upper plate was instead treated with octadecyltriethoxysilane (C18), with DMOAP in the pores.

\subsection{Cell assembly}
Two strips of \SI{50}{\micro\meter} copper foil (Thermo Fisher Scientific) served as spacers and electrodes, separated by \SI{1}{\centi\meter}. Treated substrates were assembled with functionalized surfaces facing inward, sealed with UV-curable adhesive and cured. Cell thickness, measured by optical interferometry, was about \SI{53}{\micro\meter}. Cells were filled by capillary action with 5CB (Nematel), used as received, in the isotropic phase above $T_{NI} \approx \SI{35}{\celsius}$, then cooled to room temperature. The cell rested on an insulated stage, and the electrodes were connected to a DC high-voltage power supply (Stanford Research Systems PS350).

\subsection{Colloid functionalization}
Conductive, silver-coated, monodisperse silica microspheres (Cospheric), \SI{9}{\micro\meter} in diameter, were cleaned in a 1:2 (v/v) mixture of isopropanol and water, treated with \SI{3}{wt\percent} aqueous DMOAP at \SI{80}{\celsius} for \SI{1}{\hour} to impose homeotropic anchoring, and suspended in 5CB (Nematel).

\subsection{Microscopy and tracking}
Bright-field and crossed-polarizer images were acquired on a Zeiss Axio Vert.A1 inverted microscope. Migration was recorded at \SI{20}{fps}, and defect positions were tracked with Trackpy \cite{trackpy}, which implements the Crocker-Grier algorithm \cite{Crocker1996}, to obtain instantaneous velocities. 

\subsection{Energy landscape reconstruction}
Landscapes were reconstructed from trajectories using eq~\ref{eq:landscape}, assuming overdamped motion in which the force on the defect is balanced by viscous drag, $F = C_D^* v$. Elastic landscapes were obtained from relaxation trajectories at zero field, as the defect relaxed from the pore center into a lobe, taking the lobe position as the zero of energy. The energy released along the trajectory is $\int C_D^*|v|^2\,dt$. Flexoelectric contributions were obtained by subtracting the elastic landscape from the field-on landscape, assuming equal zero-field lobe energies and a drag coefficient independent of field. We take $C_D^* = \SI{1}{\micro\newton\second\per\meter}$, consistent with $\gamma_1\ell$ for rotational viscosity $\gamma_1 \approx \SI{0.08}{\pascal\second}$ and $\ell$ of order the pore depth. 

\subsection{Landau-de Gennes simulations}
The nematic was described by the tensor order parameter $\mathbf{Q}$, with $Q_{ij} = S(n_in_j - \delta_{ij}/3)$ in the uniaxial phase \cite{Andrienko2018}. The free energy comprises phase, elastic, electric and Rapini-Papoular anchoring terms,
\begin{equation}
F[\mathbf{Q}] = \int_\Omega \left(f_\mathrm{phase} + f_\mathrm{elastic} + f_\mathrm{E}\right)dV + \int_{\partial\Omega}\frac{W}{2}\mathrm{Tr}(\mathbf{Q} - \mathbf{Q}_S)^2\,dS,
\end{equation}
with $f_\mathrm{phase} = \frac{A}{2}\mathrm{Tr}(\mathbf{Q}^2) + \frac{B}{3}\mathrm{Tr}(\mathbf{Q}^3) + \frac{C}{4}(\mathrm{Tr}\,\mathbf{Q}^2)^2$, $A = -1.72\times10^5$, $B = -2.12\times10^6$ and $C = 1.73\times10^6$~\si{\joule\per\meter\cubed}, giving $S_0 = 0.8$ in the uniaxial ground state. The three-constant elastic energy is
\begin{equation}
f_\mathrm{elastic} = \frac{L_1}{2}\partial_kQ_{ij}\partial_kQ_{ij} + \frac{L_2}{2}\partial_jQ_{ij}\partial_kQ_{ik} + \frac{L_6}{2}Q_{lk}\partial_lQ_{ij}\partial_kQ_{ij},
\end{equation}
with $L_1 = 5.5$, $L_2 = 4.1$ and $L_6 = \SI{4.8}{\pico\newton}$. The one-constant approximation retains only $L_1 = \SI{10}{\pico\newton}$. The electric energy is
\begin{equation}
f_\mathrm{E} = -\frac{1}{2}\varepsilon_0\varepsilon_a Q_{ik}E_iE_k - \mathbf{P}\cdot\mathbf{E},
\end{equation}
with $\varepsilon_a = 12$ and flexoelectric polarization including order-electric terms \cite{Blow2013},
\begin{equation}
P_i = G_1\partial_jQ_{ij} + G_2\partial_j(Q_{ik}Q_{kj}) + G_3\partial_i(Q_{kj}Q_{kj}) + G_4(Q_{ik}\partial_jQ_{kj} - Q_{kj}\partial_jQ_{ik}),
\end{equation}
with $G_i = \SI{-10}{\pico\coulomb\per\meter}$. For these values $e_1 = G(S + \tfrac{4}{3}S^2) < 0$ and $e_1 - e_3 = 2GS^2 < 0$. The simulation box measured $L_x=\SI{1.2}{\micro\meter}$, $L_y=\SI{0.7}{\micro\meter}$, and $L_z=\SI{0.5}{\micro\meter}$. An SU8 layer of thickness $h=\SI{250}{\nano\meter}$ occupied the lower half of the domain, leaving a $\SI{250}{\nano\meter}$ gap between its top surface and the upper boundary. The pillbox-shaped pore extended through the full thickness of the SU8 layer, with an overall length $a=\SI{1}{\micro\meter}$ along $x$, a width $b=\SI{0.5}{\micro\meter}$ along $y$, and two rounded end lobes, each with a radius of curvature of $\SI{250}{\nano\meter}$. Homeotropic anchoring with $W = \SI{e-2}{\joule\per\meter\squared}$ was imposed on the pore floor and walls, and the upper boundary was free ($W_\mathrm{lid} = 0$). $\mathbf{Q}$ evolved by gradient descent with rotational viscosities $\gamma = \SI{0.087}{\pascal\second}$ in the bulk and $\gamma_S = \SI{0.087e-6}{\pascal\meter\second}$ at the boundary, chosen so that the boundary response time $\gamma_S/W$ is shorter than the bulk elastic response time $\gamma\ell^2/L$ at the length scales $\ell \approx \SI{250}{\nano\meter}$ considered. These affect transients but not steady states. 

The governing nonlinear partial differential equations are implemented in dimensional form and solved using the finite-element method in COMSOL Multiphysics (Mathematics Module) through time-dependent studies. Equilibrium radial, hyperbolic, and escaped configurations are obtained from distinct initial ansatzes (see Supporting Information) using a two-stage relaxation protocol. The system is first relaxed for $\SI{100}{\milli\second}$ on a coarse mesh under the one-constant (1C) elastic approximation. The resulting configuration is then relaxed for a further $\SI{100}{\milli\second}$ on a refined mesh under either the 1C or three-constant (3C) elastic approximation. For the radial and hyperbolic configurations, the initial hedgehog is located at $(x_d,y_d,z_d)=(0.05a,0,3L_z/8)$, slightly displaced from the pore center along $x$.

The mesh consists of triangular boundary elements and tetrahedral volume elements. For the radial and hyperbolic configurations, local refinement is applied along the pore edges and along the line parallel to $\mathbf{e}_x$ at $y=0$ and $z=3L_z/8$. The height of this refinement line was chosen based on the equilibrium defect position obtained in preliminary simulations using a coarse, uniformly sized mesh. Mesh convergence tests confirmed that further refinement did not significantly alter the solution.

Switching simulations start from the equilibrium defect configurations and follow Landau–de Gennes relaxation under a constant applied electric field for $\SI{100}{\milli\second}$. All time-dependent studies use COMSOL's implicit backward differentiation formula (BDF) solver with automatic (“free”) time stepping, which adaptively selects the internal time steps.

\begin{suppinfo}
Director configurations in circular pores (Figure S1); initial conditions for Landau-de Gennes relaxation of metastable states; simulation and experimental length scales.
\end{suppinfo}

\begin{acknowledgement}
C.D. and Z.Z. were supported by the Army Research Office under grant W911NF-22-1-0205. J.-J.E. was supported by the National Science Foundation Graduate Research Fellowship Program under Grant No. DGE-1845298. Y.C. was supported on Grant No. NIH R35 GM158340. This work was carried out in part at the Singh Center for Nanotechnology at the University of Pennsylvania, which is supported by the NSF National Nanotechnology Coordinated Infrastructure Program under grant NNCI-2025608.
\end{acknowledgement}

\section*{Author Contributions}
Z.Z. and C.D. contributed equally. Z.Z. performed most of the experiments and their analysis. C.D. performed the Landau-de Gennes simulations of the lock-and-key interaction and of the flexoelectric motion of defects. C.D. and J.-J.E. also performed initial experiments and contributed to their analysis. Y.C. and A.J.P. designed and performed the molecular simulations. B.E. contributed to discussions. N.E. and K.J.S. directed the research. All authors participated in writing the manuscript.

\section*{Notes}
The authors declare no competing financial interest.

\nocite{Castles2012,Giomi2017,Link2001,OhzonoUDSO,PieranskiEPJE2019,SenyukEPAT,Sreenilayam2015}
\bibliography{refs}

\clearpage
\begin{center}
{\Large\textbf{Supporting Information}}\\[4pt]
{\large Flexoelectric Switching of Lock-and-Key Topological Bits}
\end{center}
\setcounter{section}{0}\setcounter{figure}{0}\setcounter{table}{0}\setcounter{equation}{0}
\renewcommand{\thesection}{S\arabic{section}}
\renewcommand{\thefigure}{S\arabic{figure}}
\renewcommand{\thetable}{S\arabic{table}}
\renewcommand{\theequation}{S\arabic{equation}}
\section{Director configurations in circular pores}

\begin{figure}
\centering
\includegraphics[width=0.8\textwidth]{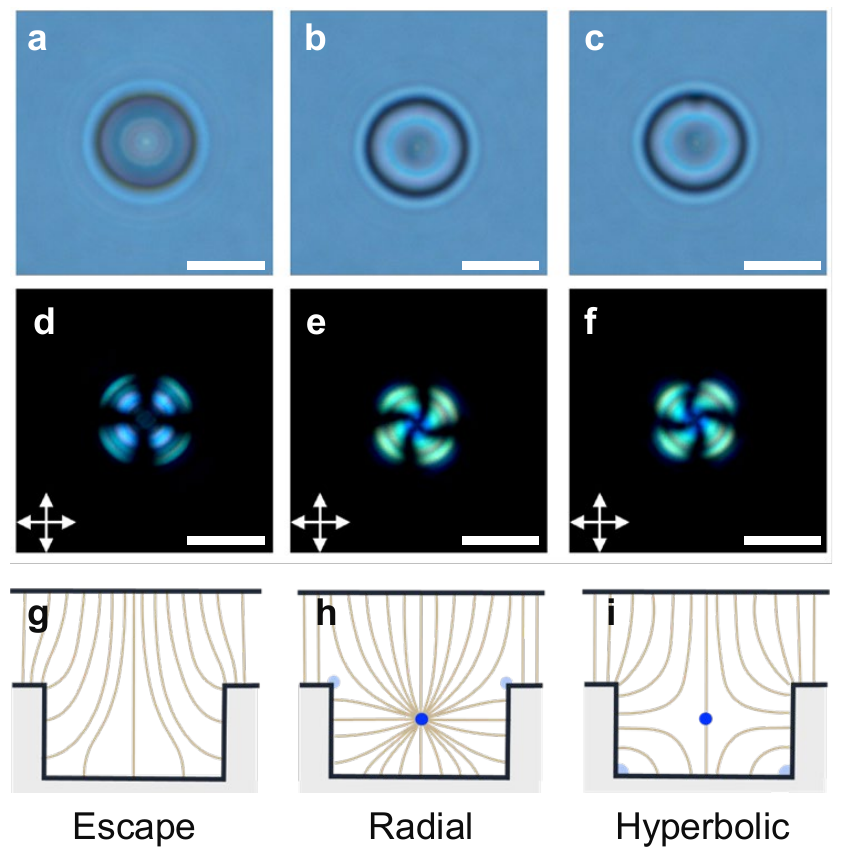}
\caption{Director configurations in circular pores, \SI{5}{\micro\meter} deep. (a--c) Bright-field and (d--f) crossed-polarizer images. The textures contain escaped and point-defect states; (b,e) and (c,f) are opposite chiral variants of the same point-defect state. (g--i) Schematic director fields for the escaped texture, a radial hedgehog and a hyperbolic hedgehog. The hyperbolic configuration is a candidate identified by modeling and is not assigned to the textures in (a--f). Scale bars, \SI{10}{\micro\meter}.}
\label{fig:circ}
\end{figure}

Three configurations are observed when 5CB fills circular pores with homeotropic walls and floor, an escaped texture without singularity and textures containing a point defect (Figure~\ref{fig:circ}). Under crossed polarizers, a homeotropic director along the optical axis appears dark, and director tilt produces birefringence and interference colors. In the escaped texture, a broad dark center indicates a homeotropic director through the pore depth, with brightness increasing toward the curved wall as the director escapes from it. In bright field, a point defect appears as a small dark spot at its core, and under crossed polarizers it shows the four-brush pattern of a defect of topological charge $\pm1$. Given their configuration in the pillbox, radial and hyperbolic hedgehogs would have identical projections on the plane transverse to the optical axis of the microscope, therefore the four-brush patterns would co-rotate with the cross-polarizers and we could not distinguish them with this method. The coexistence of textures under nominally identical conditions reflects multistability, with distinct local minima of the elastic free energy separated by barriers large compared to $k_BT$ \cite{Tran2016,Kralj1995}.

With C18 on the upper plate and DMOAP in the pores, escaped and defect-bearing textures coexisted, with about 80\% of pores defect free. With DMOAP on all surfaces, after \ce{O2} plasma activation, all pores contained a lobe defect. 

\section{Initial conditions for metastable configurations}

The Landau-de Gennes free energy is not convex and can have several local minima, particularly when the confining geometry has sharp edges \cite{SenyukEPAT}. To relax toward each candidate state, $\mathbf{Q}$ was initialized from an ansatz close to the final configuration \cite{OhzonoUDSO}, given in terms of the zenithal and azimuthal angles $\theta$ and $\phi$ of the director (Table~\ref{tab:ansatz}). Here $\phi_\mathrm{loc} = \mathrm{atan2}(y,x)$ and $\theta_\mathrm{loc} = \mathrm{atan2}(\sqrt{x^2+y^2},z)$ are measured from the initial defect position $(x_d, y_d, z_d)$. Radial and hyperbolic defects were initialized at $x_d = 0.05~a$, $y_d = 0$ and $z_d = \frac{3}{8}L_z$, and escaped textures at $x_d = y_d = 0$. 

\begin{table}
\caption{Initial director fields.}
\label{tab:ansatz}
\begin{tabular}{lcc}
\hline
state & $\theta$ & $\phi$ \\
\hline
radial & $\theta_\mathrm{loc}$ & $\phi_\mathrm{loc}$ \\
hyperbolic & $\pi - \theta_\mathrm{loc}$ & $\phi_\mathrm{loc}$ \\
escaped up & $2\arctan(\sqrt{x^2+y^2}/r_0)$ & $\phi_\mathrm{loc} + \pi$ \\
escaped down & $2\arctan(\sqrt{x^2+y^2}/r_0)$ & $\phi_\mathrm{loc}$ \\
\hline
\end{tabular}
\end{table}

\section{Simulation and experimental scales}

The Landau-de Gennes simulations are not performed on the scale of the experiments. Rather, they resolve the tensor order parameter, including the defect core, whose size is set by the nematic correlation length of order \SI{10}{\nano\meter}. Resolving this length across a domain of experimental size, tens of micrometers in each direction, is computationally prohibitive, so the simulated domain does not match the experimental pore in size. The simulations reveal important mechanistic information, including that a radial hedgehog not required by topology can be stabilized in a pore, and that this defect can be moved by flexoelectric coupling.

\end{document}